\documentclass[prc,showpacs,showkeys,superscriptaddress,nofootinbib,twocolumn,floatfix]{revtex4}
\usepackage{amsfonts}
\usepackage{graphicx,color,amsmath,amssymb}

\def\blfootnote{\xdef\@thefnmark{}\@footnotetext}

\begin{document}

\title{Non-perturbative Enhancement of Heavy Quark-Antiquark Pair Annihilation in the Quark-Gluon Plasma}
\author{Xin Wu}
\affiliation{Department of Applied Physics, Nanjing University of Science and Technology, Nanjing 210094, China}
\author{Min He}
\affiliation{Department of Applied Physics, Nanjing University of Science and Technology, Nanjing 210094, China}

\date{\today}

\begin{abstract}
The annihilation of heavy particles close to thermal equilibrium, which plays a prominent role in the chemical equilibration of heavy quarks in the Quark-Gluon Plasma (QGP) as well as in many classic dark matter scenarios, is reexamined. We derive a scattering amplitude that resums
near-threshold attractive interaction of the annihilating particles in terms of the non-relativistic scattering $T$-matrix and Green's function, thereby
capturing the pertinent Sommerfeld enhancement from both nonperturbative scattering state and bound state solutions. The derived formula is of such generality that it applies to arbitrary partial wave processes of two-particle annihilation, and enables to incorporate finite widths of the annihilating particles. In a screened potential model, the non-perturbative scattering amplitude is computed and the Sommerfeld enhancement is identified for heavy quark-antiquark annihilation in the QGP.

\end{abstract}


\maketitle

\section{Introduction}
\label{sec_intro}
Heavy particle annihilation near threshold represents a phenomenon that is encountered in different context. For example, a heavy quark ($Q$) and an anti-heavy quark $\bar{Q}$ in the Quark-Gluon Plasma (QGP) may annihilate into gluons toward chemical equilibration (although at a rate much slower than the kinetic equilibration)~\cite{Bodeker:2012gs,Bodeker:2012zm,Kim:2016zyy,Kim:2017rgb}. Reversing the direction, pair production of heavy particles has been widely studied in literature, {\it e.g.} the top-antitop quark pair production at collider energies~\cite{Appelquist:1975ya,Fadin:1990wx,Strassler:1990nw,Hoang:1999zc}. It is also of particular relevance to the dark matter search in connection with the Weakly Interacting Massive Particles (WIMP) annihilating into Standard Model particles~\cite{Hisano:2002fk,Hisano:2004ds,ArkaniHamed:2008qn}.

In all the aforementioned examples, the annihilation (or production) rate of the slowly moving particles could be much enhanced if the mutual attractive interaction prior to annihilation is taken into account, a phenomenon known as the Sommerfeld effect~\cite{Landau_QM}. Indeed, for scattering of non-relativistic particles with a light force mediator, the Feynmann amplitude of a ladder diagram picks up a factor of $\alpha/v_{\rm rel}$ (where $\alpha$ denotes the pertinent coupling and $v_{\rm rel}$ the relative velocity between the scattering particles) per ``rung". Therefore, at low $v_{\rm rel}$, these ladder diagrams are not any longer perturbative and need to be resummed to yield a non-perturbative solution.

Concerning the pair annihilation of a heavy quark and an antiquark ($Q\bar{Q}$) in near-thermal equilibrium at typical temperatures ($\sim$ a few hundred \,MeV) of the QGP created in relativistic heavy-ion collisions~\cite{Shuryak:2014zxa,Busza:2018rrf,Dong:2019byy,Rothkopf:2019ipj}, it was shown~\cite{Bodeker:2012zm,Kim:2016zyy} that the Sommerfeld enhancement factor can be defined within the non-relativistic QCD framework~\cite{Bodwin:1994jh} in terms of an imaginary-time 2-point correlator of the heavy quark singlet operators, whose spectral function has been measured on lattice~\cite{Kim:2016zyy}. A large enhancement much above the perturbative prediction in the color-singlet channel was identified, in association with the formation of the $Q\bar{Q}$ bound state that offers a less suppressed Boltzmann weight because of the binding~\cite{Kim:2016zyy}.

The present work aims to reexamine the Sommerfeld effect in the context of the heavy quark-antiquark pair annihilation in the QGP. We derive a formula for the heavy quark-antiquark scattering amplitude that resums the attractive interaction in the color-singlet channel into a $T$-matrix, from which the Sommerfeld enhancement factor from both nonperturbative scattering state and bound state solutions can be extracted. While the existing studies have usually been limited to $s$-wave scattering, we show that the derived formula applies to arbitrary partial waves and also allows to incorporate finite quark width effects. Within a screened potential model, we compute the pertinent enhancement factor as a function of energy as well as the thermally averaged enhancement that is shown to agree with the lattice results~\cite{Kim:2016zyy}.

\section{Deriving the Sommerfeld enhancement factor}
\label{sec_derivation}

\subsection{Nonperturabtive scattering amplitude: master formula}
\label{ssec_masterformula}

\begin{figure} [!t]
\includegraphics[width=1.05\columnwidth]{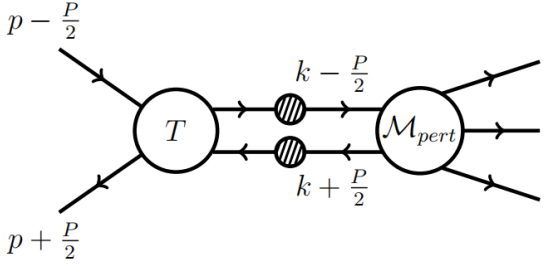}
\vspace{-0.3cm}
\caption{(Color online) Annihilation of a pair of heavy quark and antiquark, involving a nonperturbative scattering represented by the $T$-matrix prior to perturbative annihilation.}
\label{QQbar_nonperturbative_annihilation}
\end{figure}

As already pointed out in Ref.~\cite{Appelquist:1975ya}, the transition of a $Q\bar{Q}$ pair to hadronic matter can be separated into two parts: a part that contains all possible interactions between the $Q$ and $\bar{Q}$, followed by the second part representing their annihilation into gluons (and light quarks). This is depicted in Fig.~\ref{QQbar_nonperturbative_annihilation}, where the $T$-matrix resums attractive interactions to infinite order and $\mathcal{M}_{\rm pert}$ denotes the perturbative annihilation amplitude. One notes that while the effective coupling responsible for the $Q\bar{Q}$ binding prior to annihilation could be very large, the coupling strength associated with the annihilation into final states remains weak as governed by the large mass of $Q$ and $\bar{Q}$.

The full nonperturbative amplitude represented in Fig.~\ref{QQbar_nonperturbative_annihilation} is expressed as
\begin{align}\label{M_nonpert}
\mathcal{M_{\rm nonpert}}(p\rightarrow anything)=\int\frac{d^4k}{(2\pi)^4}T(k,p; P)\nonumber \\\times S(k+\frac{P}{2})S(k-\frac{P}{2})\mathcal{M_{\rm pert}}(k \rightarrow anything).
\end{align}
In non-relativistic limit, the approximation of instantaneous interaction is appropriate, which removes the dependence of the $T$-matrix (and of $\mathcal{M}_{\rm pert}$ as well) on the energy component $k^0$ of relative momentum $k$ between the intermediate $Q$ and $\bar{Q}$. As a result, upon writing the non-relativistic (scalar) $Q$ and $\bar{Q}$ propagators in the center of mass (CM) frame with total momentum $P=(2m_Q+E,\vec P=0)$ and quark width $\Gamma$~\cite{Strassler:1990nw}
\begin{align}\label{Q-Qbar_propagators}
S(k+\frac{P}{2})=\frac{1}{2m_Q(k^0+\frac{E}{2}-\frac{{\vec k}^2}{2m_Q}+i\frac{\Gamma}{2})},\nonumber \\
S(k-\frac{P}{2})=\frac{1}{2m_Q(-k^0+\frac{E}{2}-\frac{{\vec k}^2}{2m_Q}+i\frac{\Gamma}{2})},
\end{align}
the integration over $k^0$ in Eq.~(\ref{M_nonpert}) can be worked out by using complex contour techniques
\begin{align}\label{Q-Qbar_GF1}
\int\frac{dk^0}{2\pi}S(k+\frac{P}{2})S(k-\frac{P}{2})=\frac{i}{4m_Q^2}\frac{1}{E-\frac{{\vec k}^2}{m_Q}+i\Gamma}.
\end{align}
Substituting Eq.~(\ref{Q-Qbar_GF1}) into Eq.~(\ref{M_nonpert}), the nonperturbative $Q\bar{Q}$ scattering amplitude is rewritten as
\begin{align}\label{M_nonpert_new}
\mathcal{M_{\rm nonpert}}(\vec {p}\rightarrow anything)=\frac{i}{4m_Q^2}\int\frac{d^3k}{(2\pi)^3}G_0(\vec k,E+i\Gamma)\nonumber \\
\times T(\vec k,\vec p; E+i\Gamma) \mathcal{M_{\rm pert}}(\vec k \rightarrow anything),
\end{align}
where the free two-particle Green's function reads
\begin{align}\label{Q-Qbar_GF2}
G_0(\vec k,E+i\Gamma)=\frac{1}{E-\frac{{\vec k}^2}{m_Q}+i\Gamma},
\end{align}
which is the (diagonal) matrix element of the corresponding Green's operator
\begin{align}
\hat{G}_0(E+i\Gamma)=\frac{1}{E-\hat{H}_0+i\Gamma}
\end{align}
in the basis made up by the eigenstates ($|\vec k>$, $|\vec p>$) of the free Hamiltonian $\hat{H}_0$.

Eq.~(\ref{M_nonpert_new}) serves as the master formula for the $Q\bar{Q}$ nonperturbative scattering amplitude, which takes care of the pertinent
Sommerfeld enhancement from both nonperturabtive scattering state and bound state solutions of the $T$-matrix.

\subsection{Analytical enhancement factor from nonperturbative scattering state}
\label{ssec_scatteringstate}
We first show that the Sommerfeld enhancement factor from the nonperturbative scattering state solution as discussed in literature~\cite{Fadin:1990wx,ArkaniHamed:2008qn} can be derived from the master formula (Eq.~(\ref{M_nonpert_new})). To make the derivations in a more compact manner, we first introduce the $\hat{T}$-operator that satisfies the Lippmann-Schwinger equation~\cite{Taylor_Scatteringtheory}
\begin{align}\label{LSE-T}
\hat{T}(z)=\hat{V}+\hat{V}\hat{G}_0(z)\hat{T}(z)=\hat{V}+\hat{V}\hat{G}(z)\hat{V},
\end{align}
where $z=E+i\Gamma$, and the full 2-particle Green's operator
\begin{align}\label{G_operator}
\hat{G}(z)=\frac{1}{z-\hat{H}},
\end{align}
with the full Hamiltonian $\hat{H}=\hat{H}_0+\hat{V}$.
Therefore, one has
\begin{align}\label{G0-T_operators}
G_0(\vec k,z)T(\vec k,\vec p; z)=<\vec k|\hat{G}_0(z)\hat{T}(z)|\vec p>=<\vec k|\hat{G}(z)\hat{V}|\vec p>.
\end{align}

Further, introduce {\it scattering} state $|\vec{p}+>$, the eigenstate of the full Hamiltonian with the free energy $\hat{H}|\vec{p}+>=\vec p^2/2\mu|\vec{p}+>$) ($\mu=m_Q/2$ being the reduced mass), that satisfies the scattering boundary conditions~\cite{Taylor_Scatteringtheory}. The corresponding Lippmann-Schwinger equation for $|\vec{p}+>$ reads~\cite{Taylor_Scatteringtheory}
\begin{align}\label{LSE_p+}
|\vec{p}+>=|\vec{p}>+ ~\hat{G}_0(z)\hat{V}|\vec{p}+>=|\vec{p}>+ ~\hat{G}(z)\hat{V}|\vec{p}>.
\end{align}
Combining Eq.(\ref{G0-T_operators}) and Eq.(\ref{LSE_p+}), it is legitimate to rewrite
\begin{align}\label{G0TPsip}
G_0(\vec k,z)T(\vec k,\vec p; z)=<\vec k|\vec p+> \equiv \widetilde{\Psi}_{\vec p}(\vec k),
\end{align}
where we have dropped the term  $<\vec k|\vec p>$ that vanishes for typical $\vec k\neq \vec p$, and $\widetilde{\Psi}_{\vec p}(\vec k)$ is the scattering wave function in momentum representation.

Finally plugging Eq.~(\ref{G0TPsip}) into Eq.~(\ref{M_nonpert_new}), one arrives at
\begin{align}
\mathcal{M_{\rm nonpert}}(\vec {p}\rightarrow anything)=\frac{i}{4m_Q^2}\int\frac{d^3k}{(2\pi)^3}\nonumber \\
\times \widetilde{\Psi}_{\vec p}(\vec k) \mathcal{M_{\rm pert}}(\vec k \rightarrow anything).
\end{align}
For $s$-wave scattering, $\mathcal{M_{\rm pert}^{\rm (s)}}(\vec k \rightarrow anything)=const.$, so that
\begin{align}
\mathcal{M_{\rm nonpert}^{\rm (s)}}(\vec {p}\rightarrow anything)&=const.\times\int\frac{d^3k}{(2\pi)^3}\widetilde{\Psi}_{\vec p}(\vec k) \nonumber \\
&=const.\times\Psi_{\vec p}(\vec r=0),
\end{align}
and the pertinent Sommerfeld enhancement factor
\begin{align}\label{S0}
S_0(p)=\frac{|\mathcal{M_{\rm nonpert}^{\rm (s)}}(\vec {p}\rightarrow anything)|^2}{|\mathcal{M_{\rm pert}^{\rm (s)}}(\vec {p}\rightarrow anything)|^2}=|\Psi_{\vec p}(\vec r=0)|^2,
\end{align}
where constants have been absorbed such that the free plane wave function is normalized to unity at the origin point: $\psi_{\vec p}(\vec r=0)=1$~\cite{ArkaniHamed:2008qn}. It is seen from Eq.~(\ref{S0}) that the effect of the nonperturbative interaction between scattering particles prior to annihilation is to change the value of the modulus of the scattering wave function at the origin $\Psi_{\vec p}(\vec r=0)$ relative to the free plane wave value, since the $Q{\bar Q}$ annihilation takes place locally near $\vec{r}=0$. In particular, for Coulomb potential ($V(r)=-\alpha/r$) scattering, the scattering wave function can be analytically obtained~\cite{Landau_QM}, resulting in the $s$-wave Sommerfeld enhancement factor~\cite{Fadin:1990wx}
\begin{equation}\label{analyticalS0}
S_0^{\rm Coulomb}(p)=\frac{\pi/\epsilon}{1-e^{-\pi/\epsilon}},
\end{equation}
where $\epsilon=p/(\mu\alpha)$.

For $p$-wave scattering, $\mathcal{M_{\rm pert}^{\rm (p)}}(\vec k \rightarrow anything)\propto k{\rm cos}\theta_k$, so that
\begin{align}
\mathcal{M_{\rm nonpert}^{\rm (p)}}(\vec {p}\rightarrow anything)&=const.\times\int\frac{d^3k}{(2\pi)^3}\widetilde{\Psi}_{\vec p}(\vec k)k{\rm cos}\theta_k \nonumber \\
&=const.\times\frac{\partial\Psi_{\vec p}(\vec r=0)}{\partial r},
\end{align}
and the corresponding Sommerfeld enhancement factor~\cite{Cassel:2009wt}
\begin{align}\label{S1}
S_1(p)=\frac{|\mathcal{M_{\rm nonpert}^{\rm (p)}}(\vec {p}\rightarrow anything)|^2}{|\mathcal{M_{\rm pert}^{\rm (p)}}(\vec {p}\rightarrow anything)|^2}=\frac{|R_{\vec p}'(\vec r=0)|^2}{p^2},
\end{align}
where $R_{\vec p}(\vec r)$ is the radial part of the scattering wave function $\Psi_{\vec p}(\vec r)$.
To sum up, the derived master formula (Eq.~(\ref{M_nonpert_new})) for the $Q\bar{Q}$ nonperturbative scattering amplitude reproduces the $s$-wave Sommerfeld enhancement factor in literature~\cite{Fadin:1990wx,ArkaniHamed:2008qn} and also allows to derive the enhancement factor for $p$-wave scattering~\cite{Cassel:2009wt}.

\section{Sommerfeld enhancement in a screened potential model}
\label{sec_computations}
Now working with a screened potential model, we calculate the Sommerfeld enhancement from the master formula (Eq.~(\ref{M_nonpert_new})),
for the $Q{\bar Q}$ annihilation in QGP in different partial wave processes as a function of the incident energies. In particular, we compare
the contributions from the nonperturbative scattering state and bound state solutions.

\subsection{Enhancement from nonperturabtive scattering state solutions}
\label{ssec_scatteringstate}
The $T$-matrix that enters the master formula (Eq.~(\ref{M_nonpert_new})) satisfies the Lippmann-Schwinger equation in momentum representation
\begin{equation}\label{LSE}
T(\vec p\,',\vec p; E)=V(\vec p\,',\vec p) +\int d^3\vec k V(\vec p\,',\vec k)\frac{1}{E-E_{\vec k}+i\epsilon}T(\vec k,\vec p; E),
\end{equation}
where $E_{\vec k}=k^2/2\mu$ is the eigen-energy of the incident particle. This $T$-matrix equation can be recast into an $1D$ partial wave integral equation~\cite{Taylor_Scatteringtheory},
\begin{equation}\label{partialwaveLSE}
T_l(p\,',p; E)=V_l( p\,',p) +\frac{2}{\pi}\int_0^\infty k^2dk \frac{V_l(p\,',k)}{E-E_{\vec k}+i\epsilon}T_l(k, p; E),
\end{equation}
with the partial wave potential ($j_l$ being the spherical Bessel function)
\begin{equation}\label{partialwaveV_l}
V_l(p\,',p)=\int_0^\infty r^2drj_l(p\,'r)V(r)j_l(pr).
\end{equation}

\begin{figure} [!t]
\includegraphics[width=1.05\columnwidth]{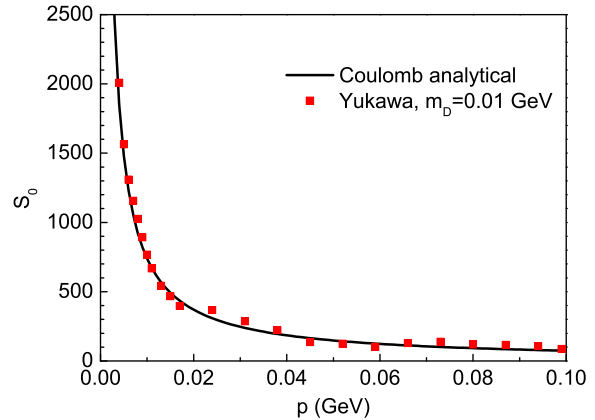}
\vspace{-0.3cm}
\caption{(Color online) The $s$-wave Sommerfeld enhancement computed from $T$-matrix scattering state solution with a Yukawa potential with coupling constant $\alpha=0.47$ and screening mass $m_D=0.01$\,GeV, in comparison with the corresponding analytical result (Eq.~(\ref{analyticalS0})) from a Coulomb potential.}
\label{S0_analytical-vs-numerical}
\end{figure}

To verify the correctness of the derived master formula, we numerically evaluate Eq.~(\ref{M_nonpert_new}) for the $s$-wave scattering state solution
and compare it to the corresponding analytical result Eq.~(\ref{analyticalS0}) (for zero quark widths). For this purpose, we work with a Yukawa potential
\begin{equation}\label{Yukawa}
V(r)=\frac{\alpha}{r}e^{-m_Dr},
\end{equation}
with coupling constant $\alpha=0.47$~\cite{Karsch:1987pv}, quark mass $m_Q=5$\,GeV ($\mu=2.5$\,GeV), and Debye screening mass $m_D=0.01$\,GeV. The tiny value of the Debye mass makes the Yukawa potential a reasonable proxy for the Coulomb potential, while still maintaining numerical convergence. Eq.~(\ref{partialwaveLSE}) is solved via the matrix inversion method and the calculated $s$-wave Sommerfeld enhancement factor is displayed in Fig.~\ref{S0_analytical-vs-numerical}. A considerable enhancement is seen in the calculated factor toward low energies, in good agreement with the analytical counterpart (Eq.~(\ref{analyticalS0})), whereas at high energies it tends to the baseline unity.

\begin{figure} [!t]
\includegraphics[width=1.05\columnwidth]{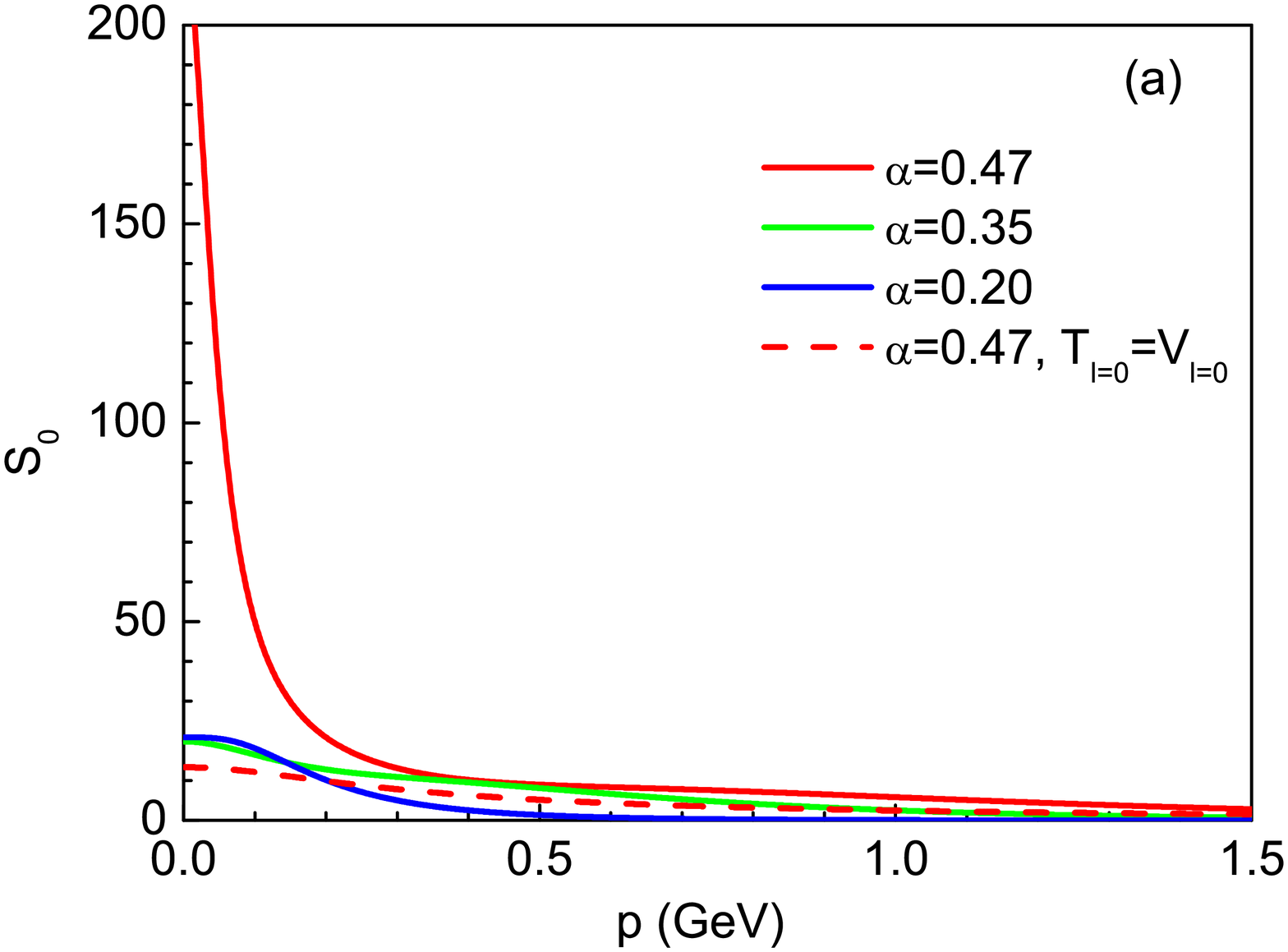}
\vspace{-0.3cm}
\includegraphics[width=1.05\columnwidth]{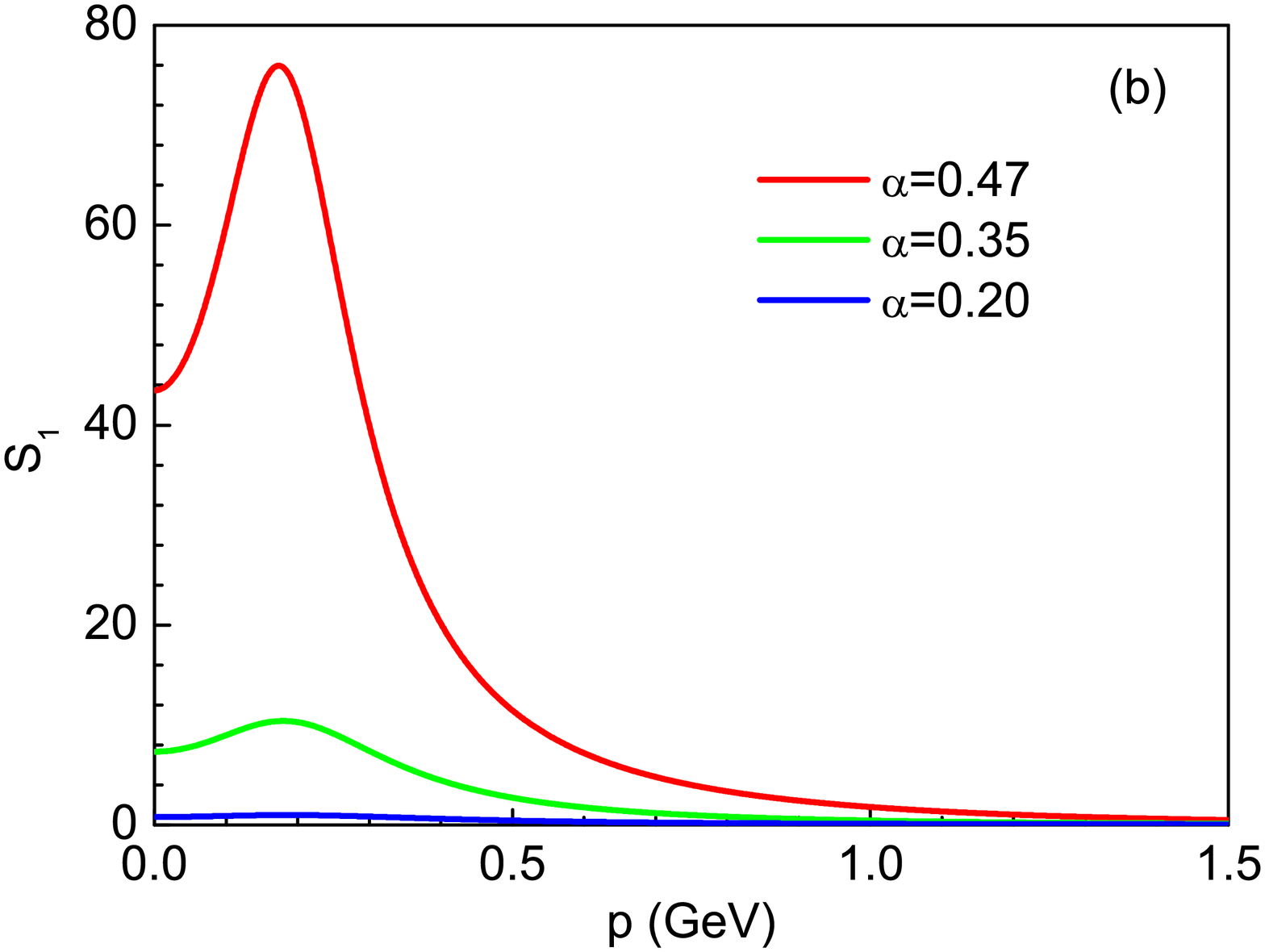}
\vspace{-0.3cm}
\caption{(Color online) The (a) $s$-wave and (b) $p$-wave Sommerfeld enhancement factors computed from $T$-matrix scattering state solutions with an input Yukawa potential with screening mass $m_D=0.4$\,GeV and zero quark widths, comparing cases with different coupling constants. The dashed line in panel (a) refers to the result without $T$-matrix resummation.}
\label{S0_S1_comparing-alpha}
\end{figure}

Now still working with such a Yukawa potential as input of the $T$-matrix but with $m_D=0.4$\,GeV (to ensure better numerical convergence), we calculate the $s$-wave and $p$-wave Sommerfeld enhancement factors for different coupling constant $\alpha=0.47$, $0.35$ and $0.20$, as displayed in Fig.~\ref{S0_S1_comparing-alpha}. One sees that for both $s$-wave and $p$-wave, a large enhancement occurs at low energies, which becomes stronger for larger coupling constant, while approaching the baseline unity toward high energies. On the other hand, if the $T$-matrix resummation is not to be done ({\it i.e.}, setting $T_{l=0}=V_{l=0}$ and removing the nonperturbative treatment), then the enhancement is much diminished, as shown by the dashed line for $\alpha=0.47$ in the upper panel of Fig.~\ref{S0_S1_comparing-alpha}. Another observation is that the $s$-wave enhancement is significantly stronger than the $p$-wave counterpart toward zero energy for the same coupling, since the isotropic $s$-wave scattering dominates the low energy regime.

\subsection{Enhancement from nonperturabtive bound state solutions}
\label{ssec_boundstate}
\begin{figure} [!t]
\includegraphics[width=1.05\columnwidth]{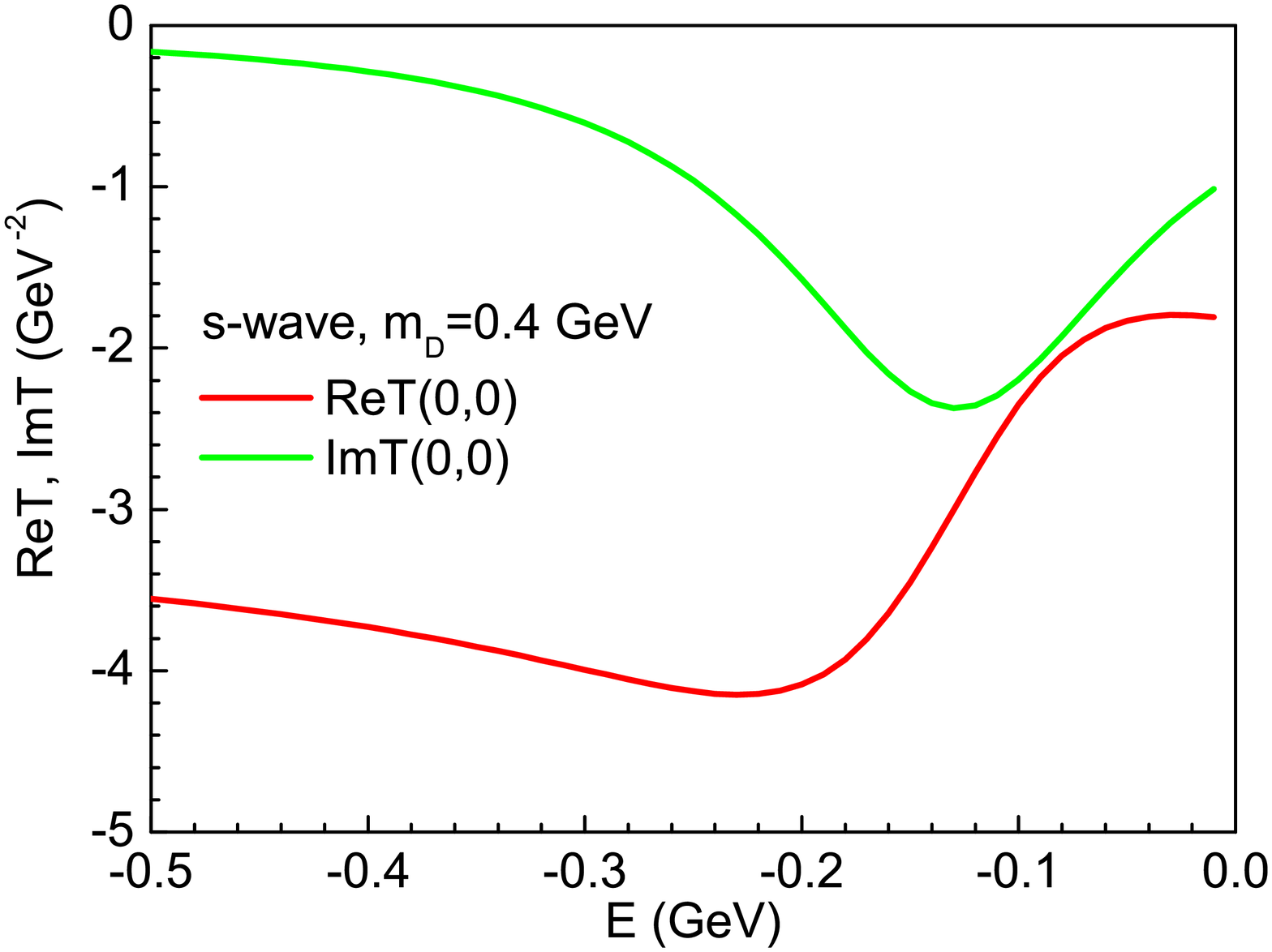}
\vspace{-0.3cm}
\caption{(Color online) The real and imaginary part of the $s$-wave $T$-matrix at vanishing relative momenta $p=p'=0$ as a function of energy computed from an input Yukawa potential with coupling constant $\alpha=0.47$, screening mass $m_D=0.4$\,GeV and a quark width $\Gamma=0.1$\,GeV.}
\label{boundstateReTImTvsE}
\end{figure}

The large enhancement of the $Q\bar{Q}$ annihilation rate as seen from lattice measurements in Ref.~\cite{Kim:2016zyy} was suggested to originate from bound state  formation and subsequent decay. To identify such a contribution in our approach, we first scrutinize the bound state solution of the $s$-wave $T$-matrix equation~(Eq.(\ref{partialwaveLSE}) with $l=0$) with $\alpha=0.47$, $m_D=0.4$\,GeV and a finite quark width $\Gamma=0.1$\,GeV still in the Yukawa potential~(Eq.(\ref{Yukawa})). A scan of the resulting $T$-matrix at vanishing relative momenta ({\it i.e.} $T_{l=0}(p=0,p'=0)$) with respect to energy yields a bound state at binding energy $E=-0.13$\,GeV, {\it i.e.}, the location of the minimum (inflection) point of the imaginary (real) part of the $T$-matrix as shown in Fig.~\ref{boundstateReTImTvsE}.

\begin{figure} [!t]
\includegraphics[width=1.05\columnwidth]{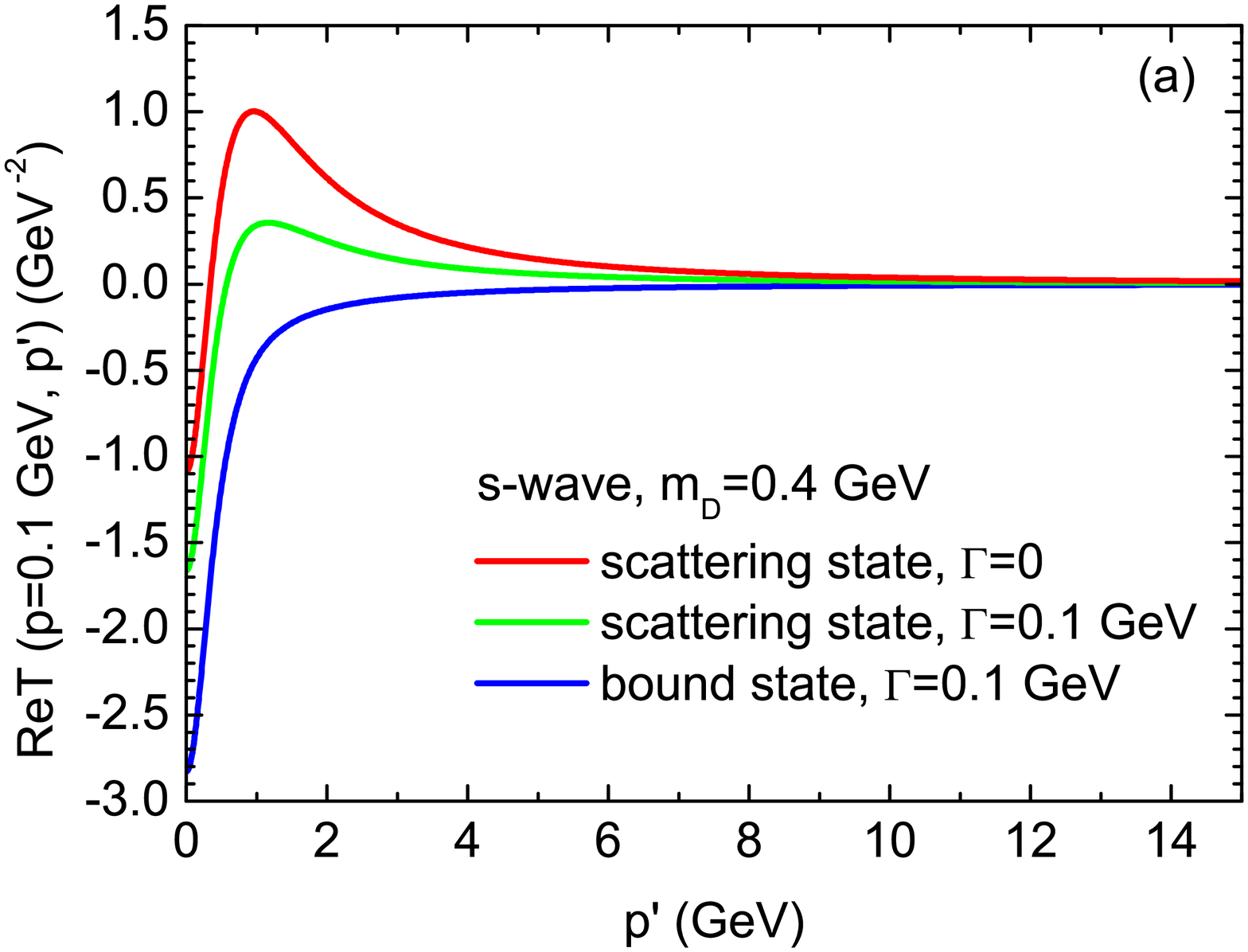}
\vspace{-0.3cm}
\includegraphics[width=1.05\columnwidth]{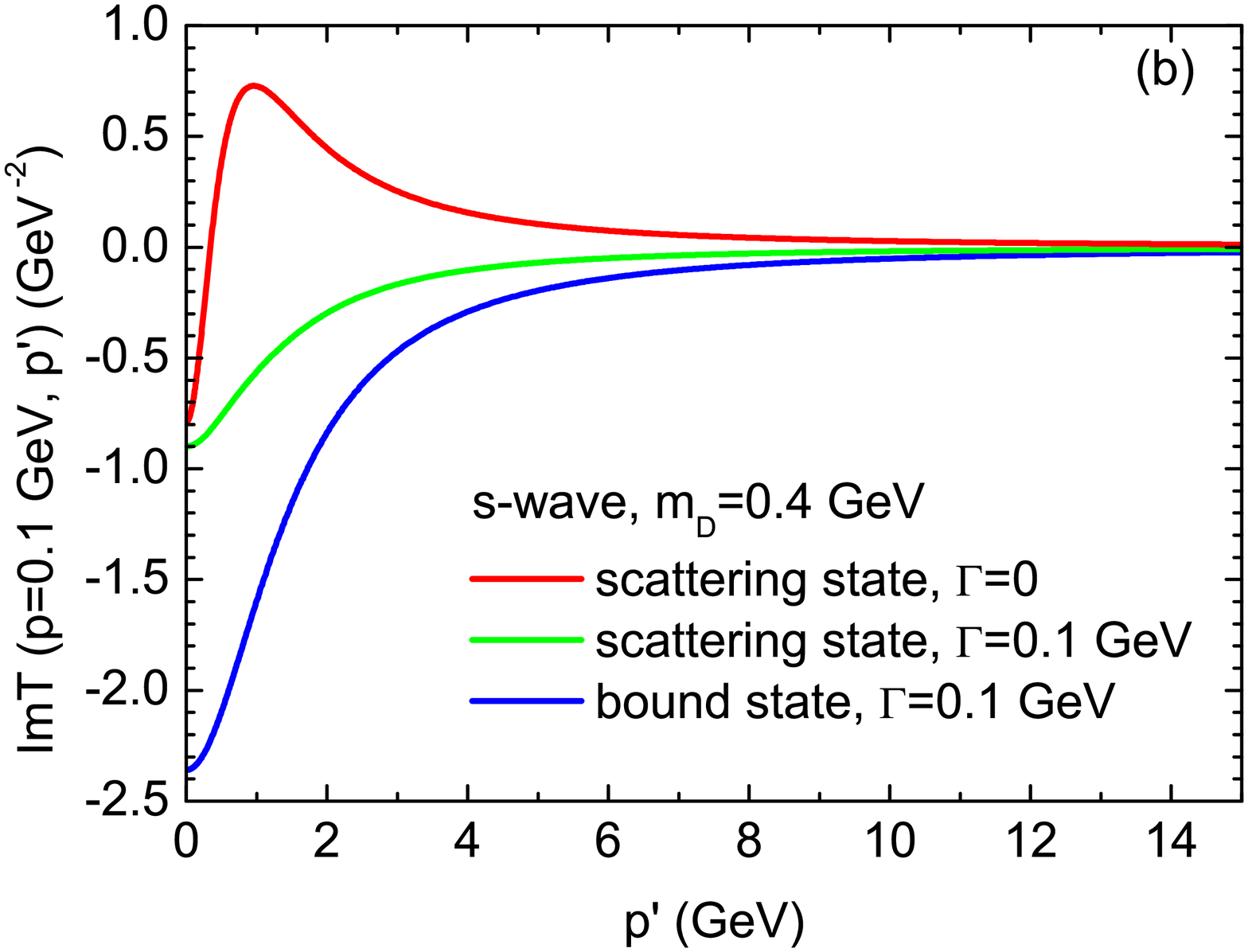}
\vspace{-0.3cm}
\caption{(Color online) The (a) real and (b) imaginary part of the $s$-wave $T$-matrix at $p=0.1$\, GeV for bound state and scattering state solutions computed from an input Yukawa potential with coupling constant $\alpha=0.47$, screening mass $m_D=0.4$\,GeV and a quark width $\Gamma=0.1$\,GeV. The results for the scattering state with zero quark width are also shown.}
\label{ReTImTvspprime_boundvsscatteringstate}
\end{figure}

Then the real and imaginary part of the bound state $T$-matrix at fixed initial state relative momentum $p=0.1$\,GeV are displayed in Fig.~\ref{ReTImTvspprime_boundvsscatteringstate} as a function of the final state relative momentum $p'$, in comparison with the corresponding scattering state amplitudes obtained with the same parameters (or up to a different quark width). Both the real and imaginary part of the $T$-matrix from the bound state solution exhibit a large amplification at low momenta relative to the scattering state amplitudes, while they all converge to zero at high momenta. Another observation is that incorporating a finite quark width $\Gamma=0.1$\,GeV suppresses the scattering state amplitudes.

\begin{figure} [!t]
\includegraphics[width=1.05\columnwidth]{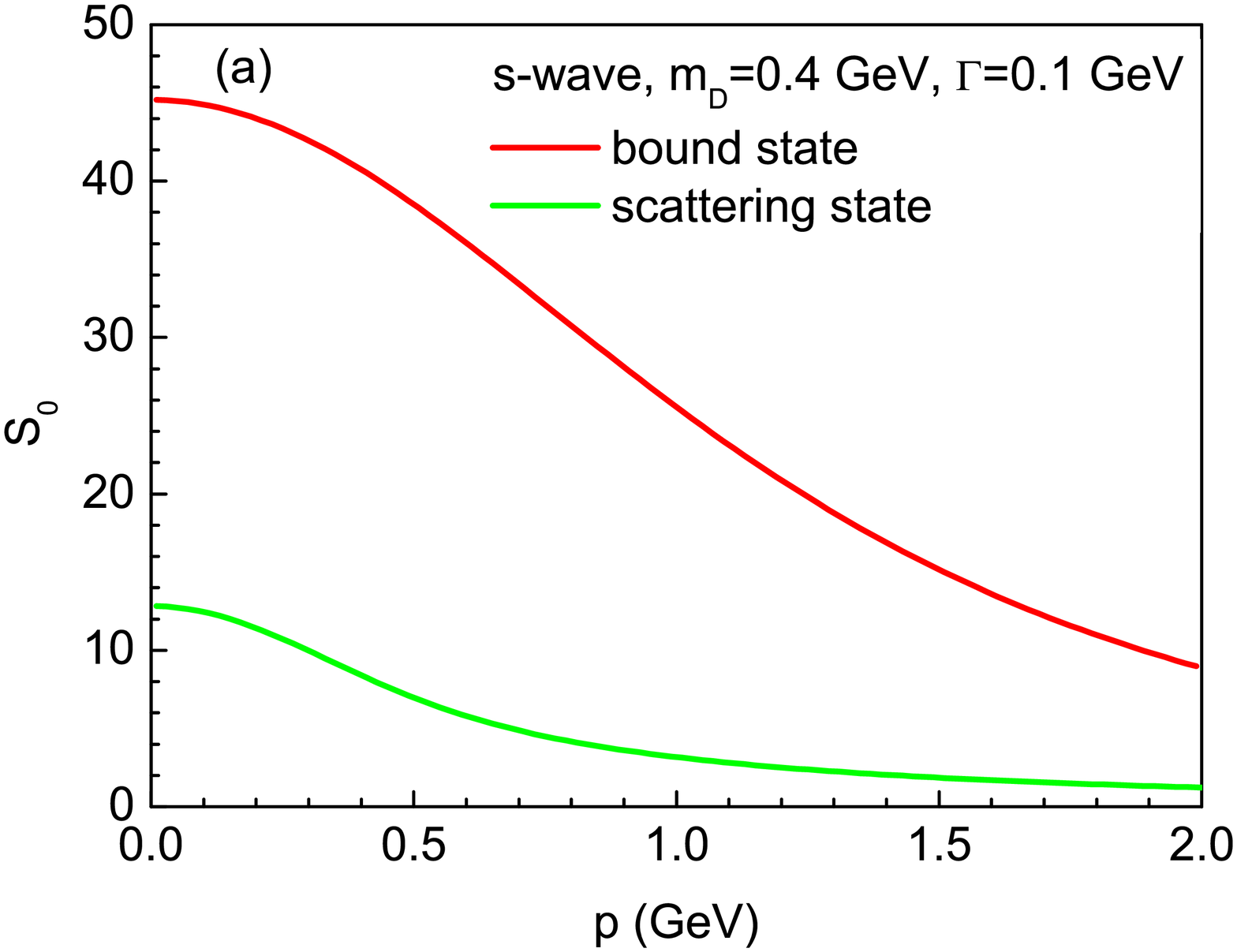}
\vspace{-0.3cm}
\includegraphics[width=1.05\columnwidth]{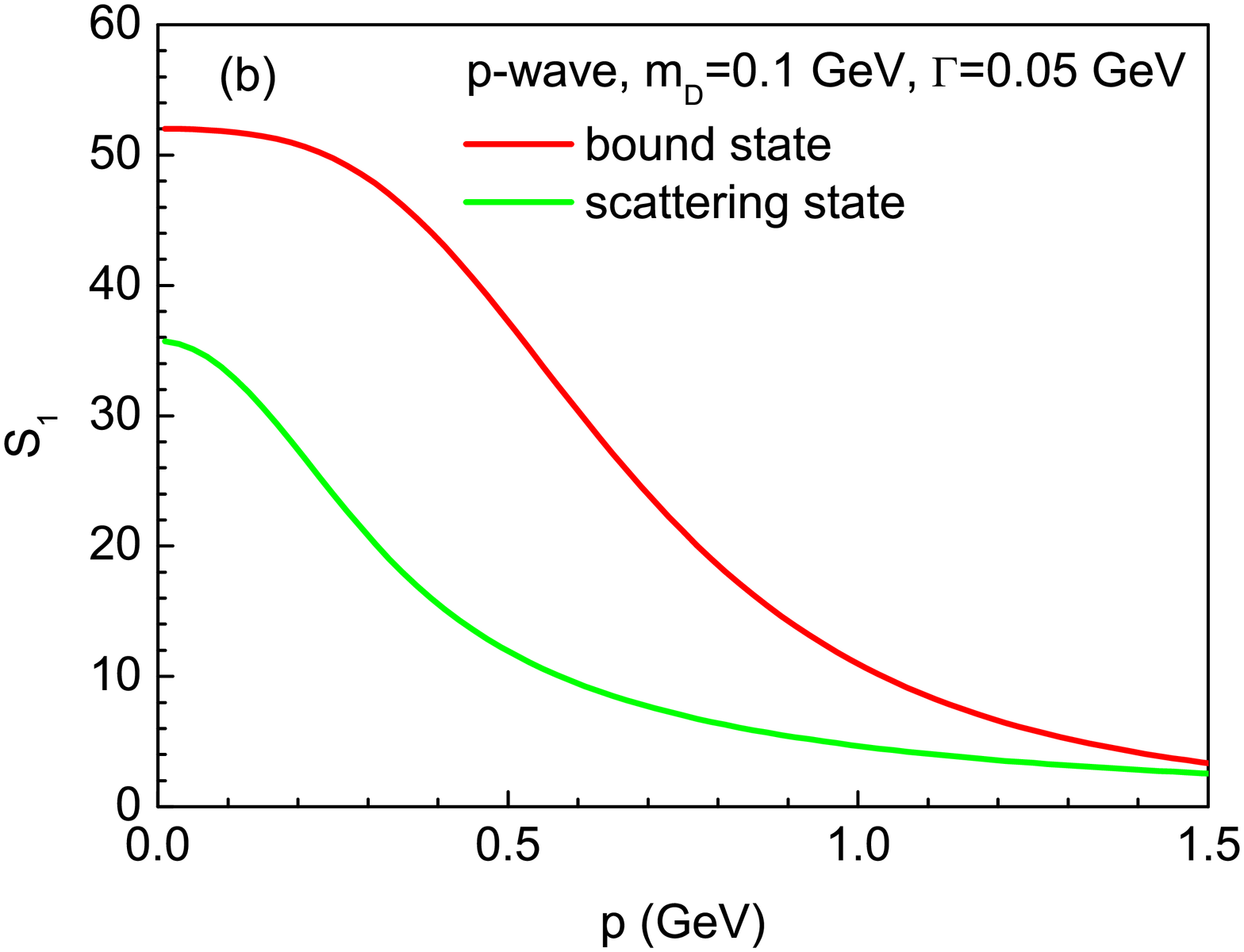}
\vspace{-0.3cm}
\caption{(Color online) The (a) $s$-wave ($\alpha=0.47$, $m_D=0.4$\, GeV and $\Gamma=0.1$\,GeV)  and (b) $p$-wave ($\alpha=0.47$, $m_D=0.1$\, GeV and $\Gamma=0.05$\,GeV) Sommerfeld enhancement factors, comparing bound state and scattering state solutions.}
\label{S0S1-vs-p_boundvsscatteringstate}
\end{figure}

Now we are in a position to use these solved $T$-matrix amplitudes to compute the pertinent Sommerfeld enhancement factor utilizing the master formula (Eq.~(\ref{M_nonpert_new})). The results for the $s$-wave enhancement factor $S_0$ as a function of the incident energy are shown in the upper panel of Fig.~\ref{S0S1-vs-p_boundvsscatteringstate}. One observes that the $S_0$ computed from the bound state solution is much magnified at all momenta relative to that from the scattering state solution (both with finite quark width $\Gamma=0.1$\,GeV, which significantly reduces the scattering state enhancement factor relative to the zero quark case shown in the upper panel of Fig.~\ref{S0_S1_comparing-alpha} with otherwise same parameters, owing to the corresponding suppression of the $T$-matrix amplitudes in Fig.~\ref{ReTImTvspprime_boundvsscatteringstate}). Also the former displays a much slower drop-off than the scattering state counterpart toward high energies where $S_0$ should converge to the baseline unity. In the $p$-wave case shown in the lower panel of Fig.~\ref{S0S1-vs-p_boundvsscatteringstate}, where we choose to use a reduced screening mass $m_D=0.1$\,GeV in the potential in order to still retain a shallow bound state solution with binding energy $E=-0.01$\,GeV, the relative enhancement of $S_1$ from the scattering state to the bound state becomes less pronounced, and both contributions tend to the baseline unity toward the same $p\geq 1.5$\,GeV.

\section{Thermally averaged Sommerfeld enhancement factor}
\label{ssec_thermalaverage}
To examine the temperature dependence, a Boltzmann factor weighted average is performed over the energy-dependent $s$-wave Sommerfeld enhancement factor~\cite{Kim:2016zyy}
\begin{equation}
<S_0>=\frac{\int d^3p~e^{-E_{\rm rel}/T}S_0(p)}{\int d^3p~e^{-E_{\rm rel}/T}},
\end{equation}
where $E_{\rm rel}$ is the energy of the relative motion of the annihilating particles.

\begin{figure} [!t]
\includegraphics[width=1.05\columnwidth]{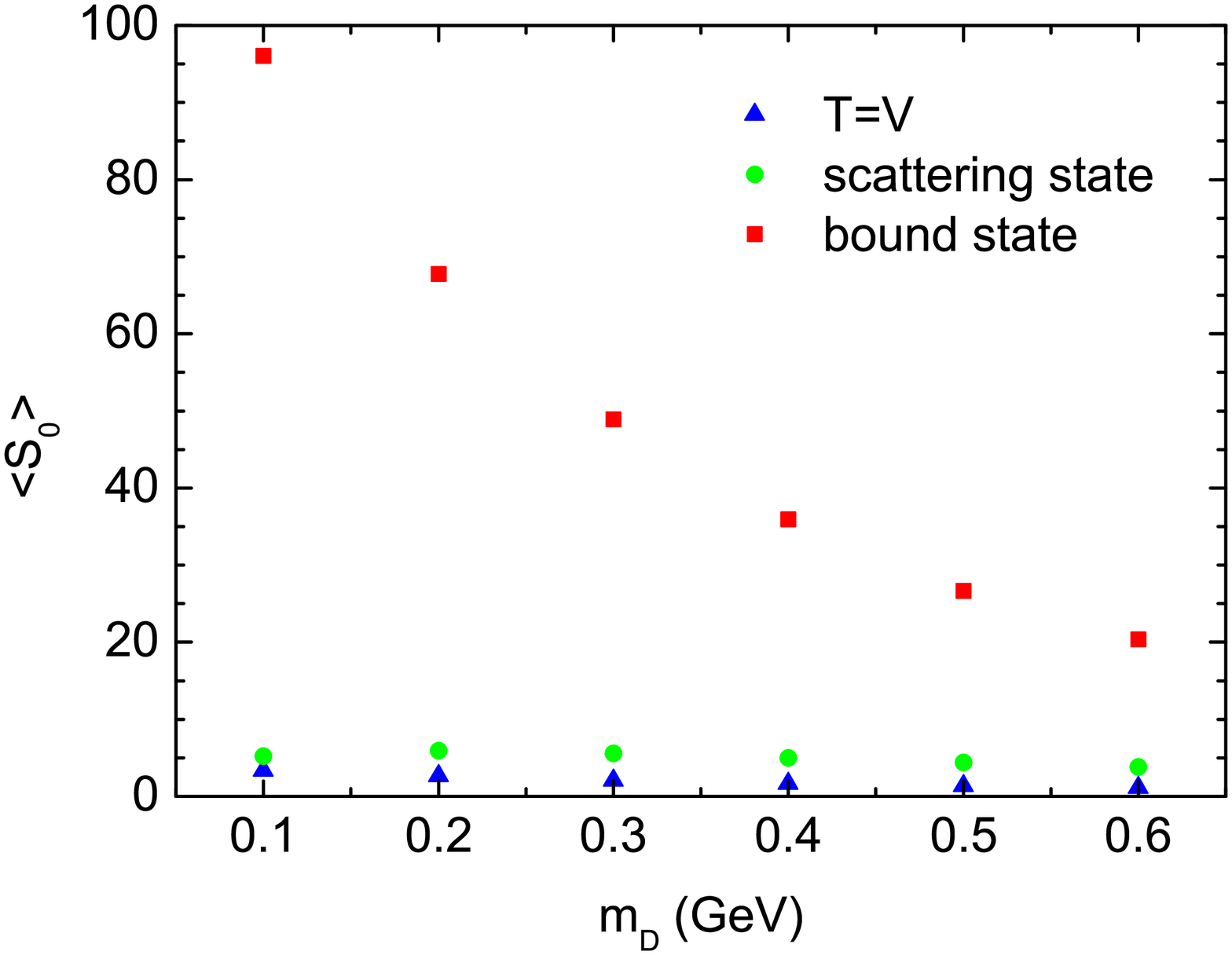}
\vspace{-0.3cm}
\includegraphics[width=0.80\columnwidth]{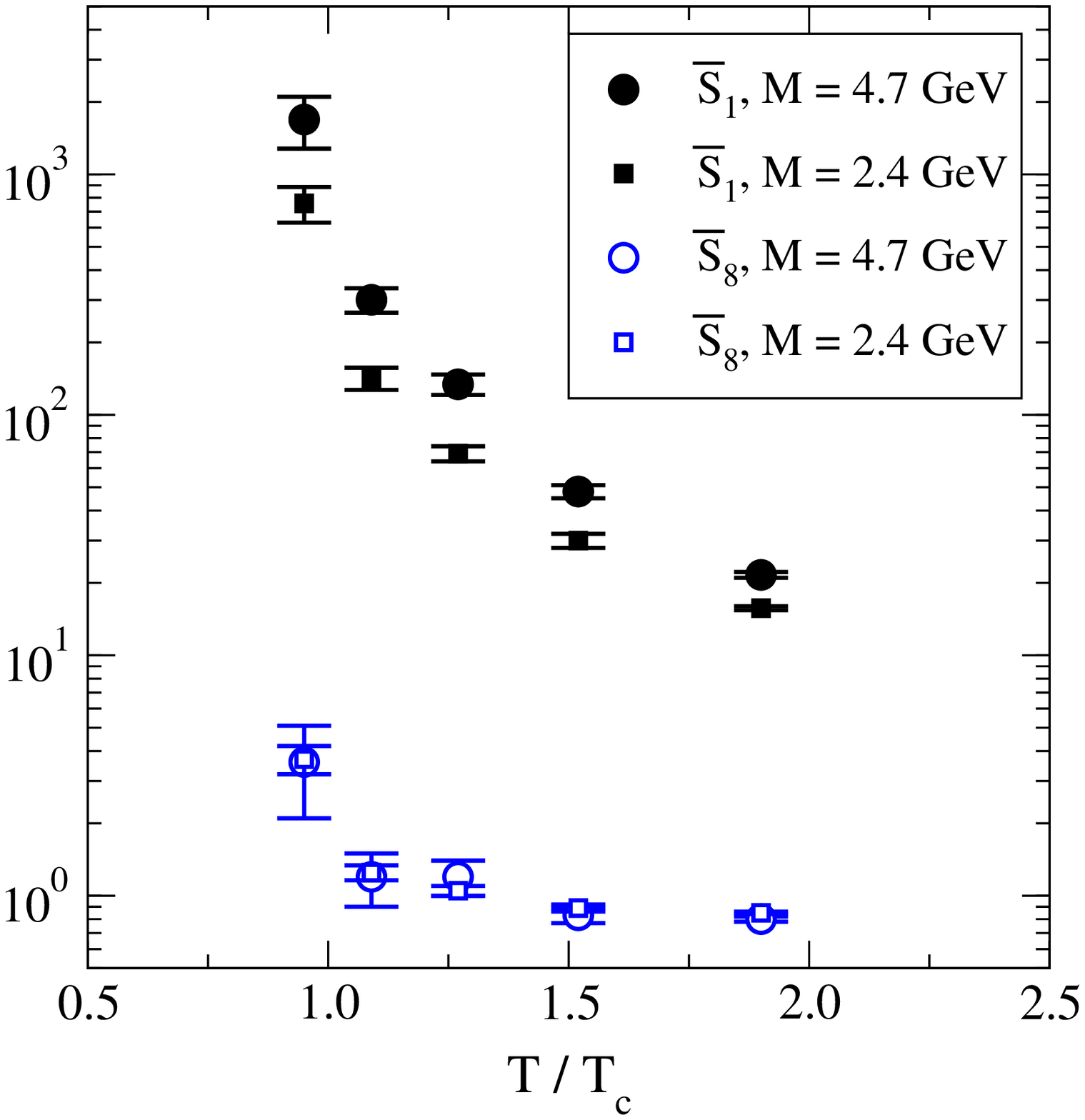}
\vspace{-0.2cm}
\caption{(Color online) Upper: the thermally averaged $s$-wave Sommerfeld enhancement factor computed from $T$-matrix with an input Yukawa potential with $\alpha=0.47$ and $\Gamma=0.1$\,GeV for varying screening masses, comparing perturbative $T=V$ scenario, nonperturbative scattering state and bound state solutions. Lower: the same quantity computed from lattice NRQCD for color-singlet ($\bar{S}_1$) and octet ($\bar{S}_8$), as a function of reduced temperatures $T/T_c$; adapted from Ref.~\cite{Kim:2016zyy}.}
\label{thermally-averaged-S0}
\end{figure}

Quoting a schematic relation between the screening mass and temperature, $m_D\sim T$, we evaluate this thermally averaged Sommerfeld enhancement factor
for both the scattering state and bound state solutions. As shown in Fig.~\ref{thermally-averaged-S0}, nonperturabtive interaction forming the scattering state only mildly enhances the $<S_0>$ relative to the perturbative evaluation ({\it i.e.}, setting $T=V$ without doing the $T$-matrix resummation). In contrast, the bound state solution results in a considerable enhancement, rendering it the dominant contribution and agreeing with the findings in lattice measurements~\cite{Kim:2016zyy}. Comparing it to the result obtained from lattice QCD estimates~\cite{Kim:2016zyy} in the color-singlet channel, a common temperature-dependent behavior is found from our computation that the averaged enhancement factor is boosted by one to two orders of magnitude toward low temperatures due to the weakening screening of the attractive $Q\bar{Q}$ potential.

\section{Summary}
\label{sec_sum}

In this work, we have reexamined the nonperturbative enhancement in the pair annihilation of nonrelativistic particles that is of direct relevance for determining the abundance of heavy weakly interacting dark matter particles~\cite{Hisano:2002fk,Hisano:2004ds,ArkaniHamed:2008qn}. We have derived a master formula (Eq.~(\ref{M_nonpert_new})) in terms of the nonrelativistic Green's function and the elastic scattering $T$-matrix that accounts for the nonperturbative interaction of the annihilating particles before the final state inelastic process and takes care of the pertinent Sommerfeld enhancement from both nonperturabtive scattering state and bound state solutions. This formula is shown to be able to recover the analytical result for the $s$-wave Coulomb scattering state and further allows us to compute the Sommerfeld enhancement factor in different partial waves and to incorporate the finite quark width effects.

Using a screened $Q\bar{Q}$ potential model as illustration, we have quantified the nonperturbative enhancement toward low energies for both scattering state and bound state solutions. We have also evaluated the thermally averaged enhancement factor which turns out to agree qualitatively with the lattice QCD estimates upon incorporating the dominant bound state contribution. The direct consequence of this enhancement is to boost the heavy quark chemical equilibration~\cite{Kim:2016zyy} in relativistic heavy ion collisions. Although heavy quark chemical equilibration is unlikely in currently available heavy ion collision experiments at RHIC and the LHC where the QGP formed attains a highest temperature $\sim 400-600$\, MeV and a life time $\leq 10$\,fm, it may be realized to a large extent in the heavy ion program at the Future Circular Collider (FCC)~\cite{Dainese:2016gch}.

\acknowledgments This work was supported by NSFC grant 11675079.

\end{document}